\documentclass[review]{elsarticle}

\journal{Journal of \LaTeX\ Templates}

\usepackage{lineno,hyperref}
\modulolinenumbers[5]
\usepackage{color}
\usepackage{ccaption}
\usepackage{booktabs}
\usepackage{float}
\usepackage{fancyhdr}
\usepackage{caption}
\usepackage{xcolor,stfloats}
\usepackage{comment}

\begin{document}

\begin{frontmatter}

\title{The Optimization Landscape of Hybrid Quantum-Classical Algorithms: from Quantum Control to NISQ Applications \tnoteref{mytitlenote}}
\tnotetext[mytitlenote]{RBW acknowledges the support by the National Key R\&D Program of China (Grants No. 2018YFA0306703 and No. 2017YFA0304304), NSFC (Grants No. 62173201 and No. 61833010). HR acknowledges the support from the US Department of Energy (DE-FG02-02ER15344) for control landscape analysis and the US ARO (W911NF-19-1-0382) for quantum information science analysis.}

\author[mymainaddress,mysecondaryaddress]{Xiaozhen Ge}
\author[mymainaddress]{Re-Bing Wu\corref{mycorrespondingauthor}}
\address[mymainaddress]{Center for Intelligent and Networked Systems, Department of Automation, Tsinghua University, Beijing, 100084, China}
\cortext[mycorrespondingauthor]{Corresponding authors}
\ead{rbwu@tsinghua.edu.cn}
\address[mysecondaryaddress]{Department of Applied Mathematics, The Hong Kong Polytechnic University, Hong Kong, China}

\author[mythirdaddress]{Herschel Rabitz}
\address[mythirdaddress]{Department of Chemistry, Princeton University, Princeton, NJ 08544, USA}


\begin{abstract}
This review investigates the landscapes of hybrid quantum-classical optimization algorithms that are prevalent in many rapidly developing quantum technologies, where the objective function is computed by either a natural quantum system or an engineered quantum ansatz, but the optimizer is classical. In any particular case, the nature of the underlying control landscape is fundamentally important for systematic optimization of the objective. In early studies on the optimal control of few-body dynamics, the optimizer could take full control of the relatively low-dimensional quantum systems to be manipulated. Stepping into the noisy intermediate-scale quantum (NISQ) era, the experimentally growing computational power of the ansatz expressed as quantum hardware may bring quantum advantage over classical computers, but the classical optimizer is often limited by the available control resources. Across these different scales, we will show that the landscape's geometry experiences morphological changes from favorable trap-free landscapes to easily trapping rugged landscapes, and eventually to barren-plateau landscapes on which the optimizer can hardly move. This unified view provides the basis for understanding classes of systems that may be readily controlled out to those with special consideration, including the difficulties and potential advantages of NISQ technologies, as well as seeking possible ways to escape traps or plateaus, in particular circumstances.
\end{abstract}


\end{frontmatter}







\section{Introduction}
\label{Sec:Introduction}
Optimization is ubiquitous in the pursuit of quantum technologies for molecular transformations~\cite{Magann2021}, high-precision sensing~\cite{Degen2017,Poggiali2018}, secure communications~\cite{Waks2002}, revolutionary computing~\cite{Resch2019}, etc. At the physical level, the classical electromagnetic fields for controlling quantum dynamics need to be optimized for improving performance~\cite{Brif2010,Glaser2015}. The past two decades have witnessed a large number of successes in quantum control, and new applications are still emerging. In a different but closely related area, variational algorithms operating on quantum circuits~\cite{Cerezo2021} also involve optimization of classical control parameters, which are expected to achieve computational advantage on current Noisy Intermediate-Scale Quantum (NISQ) devices~\cite{Preskill2018} with applications in many areas including  {quantum simulation, combinatorial optimization, quantum chemistry and quantum machine learning~\cite{Lubasch2020,Bharti2021,Benedetti2021,Plekhanov2022,Amaro2022,Dou2022}.}
\begin{figure}
	\centering
	\includegraphics[width=1\columnwidth]{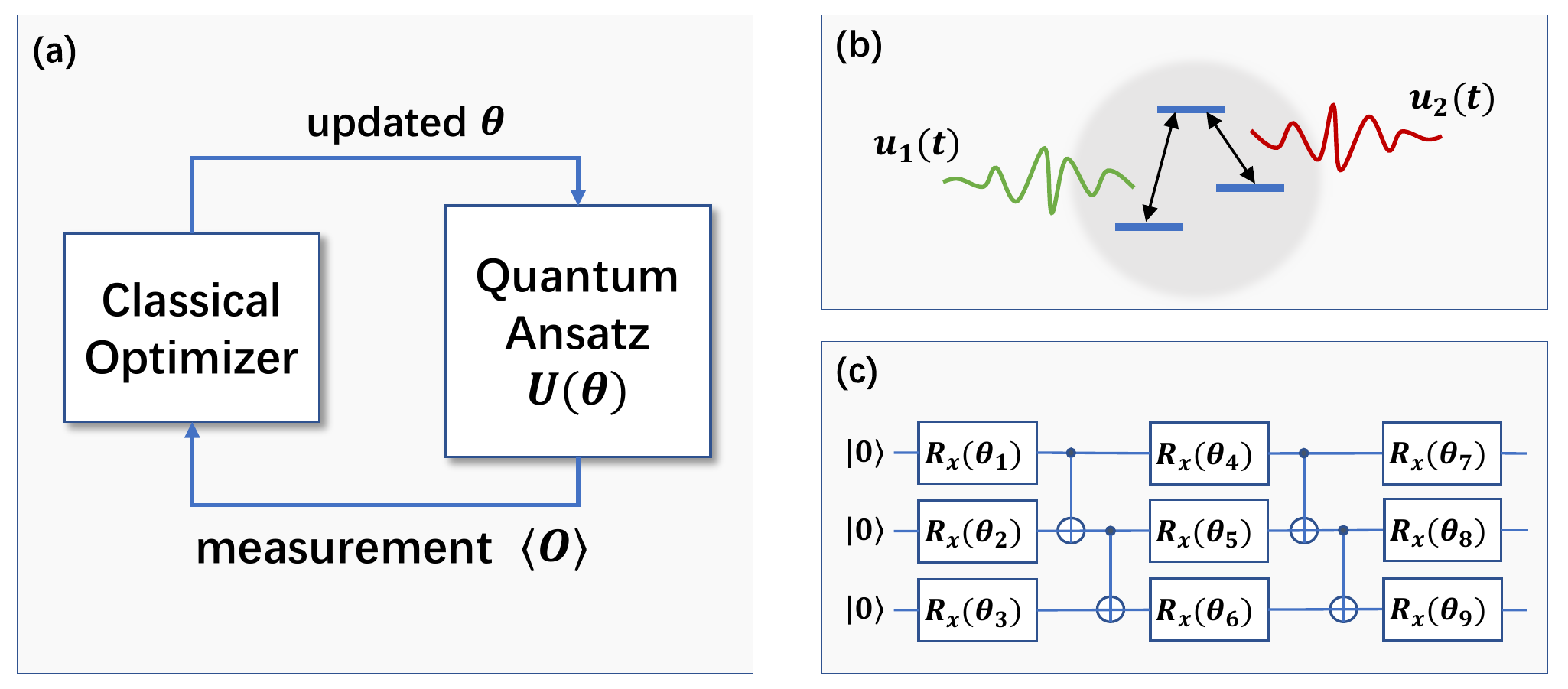}
	\caption{(a) The general setup of hybrid quantum-classical optimization system; (b) the optimization of classical fields in the control of a molecule or other modest sized quantum system; (c) the training of variational quantum circuits using a classical optimizer.}
	\label{fig:schematics}
\end{figure}

Quantum optimal control (QOC) and the variational quantum algorithm (VQA), as shown in Figs.~\ref{fig:schematics}(b) and (c), respectively, can be categorized as hybrid quantum-classical optimization algorithms in Fig.~\ref{fig:schematics}(a). Both QOC and VQA settings fit the framework in Fig.~\ref{fig:schematics}(a) which can be understood as special cases of an early paradigm~\cite{Judson1992}, differing in whether the quantum system is natural (e.g., a molecule) or engineered (e.g., coupled qubits). The objective function to be optimized is evaluated by a quantum ansatz $U(\theta)$, but the optimizer for updating the control variables $\theta$ is classical. Here we use the word ansatz to generally encompass natural or engineered quantum systems described by their respective unitary evolution operator $U$. Since greedy algorithms (e.g., gradient-based algorithms) are a frequent choice, it is fundamental to investigate whether the optimization possesses a nice landscape, i.e., whether the designed algorithm can successfully and efficiently reach favored solutions, with the landscape being the physical objective as a function of the control. Otherwise, a bad landscape (e.g., with many traps) may nullify the prospect of achieving optimal quantum performance.

Suppose that the quantum ansatz is defined on an $N$-dimensional Hilbert space, and the real variables $\theta$ to be optimized are in an $M$-dimensional space $\mathcal{X}$. In most existing quantum control applications, the number of control parameters is much larger than the effective dimension of the system, i.e., $M\gg N$; creating an arbitrary unitary $U$ generally require $M$ suitable controls satisfying $M\geq N^2$. By contrast, the opposite situation $M\ll N$ may be encountered in NISQ applications as the exponentially increasing dimensionality of the engineered quantum system grows so fast that it will soon be far greater than the number of practically tunable circuit parameters. The discussions in this review will show that the optimization landscapes of hybrid quantum-classical algorithms will experience morphological transitions when the size of quantum systems grows from small to large with respect to the available resources of the classical optimizer.

\begin{figure}
	\centering
	\includegraphics[width=1\columnwidth]{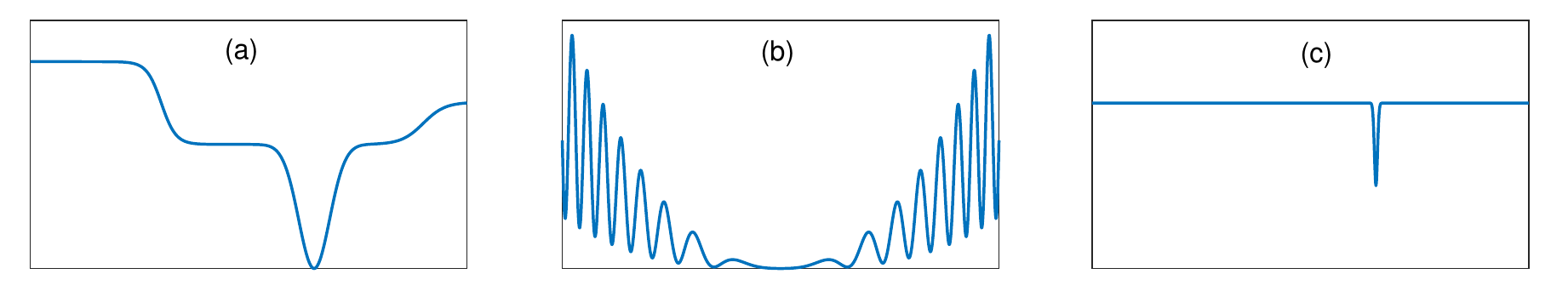}
	\caption{General forms of the landscapes: (a) a trap-free landscape with non-trapping saddles; (b) a rugged landscape with many false traps; (c) a landscape with a barren plateau.}
	\label{fig:schematics_form}
\end{figure}

The early investigation of quantum control revealed that the landscape is almost always devoid of traps when the control fields are unlimited~\cite{Rabitz2004}, as is schematically shown in Fig.~\ref{fig:schematics_form}(a). There are additional saddle points on the landscape that may slow down the optimization, but they will not halt even a greedy algorithm search for global optimal solutions~\cite{Riviello2017}. When the control resources are insufficient, e.g., when the control pulses have very limited time duration and bandwidth, false traps will emerge to likely halt the optimization procedure in a suboptimal minimum (Fig.~\ref{fig:schematics_form}(b)). Stochastic algorithms can be effective in this circumstance, but they are often costly to run. Recently, it was discovered that barren plateaus~\cite{McClean2018}, which refer to exponentially vanishing gradients with the growth of the qubit number as a consequence of the measure concentration, may become dominant on the optimization landscapes (see Fig.~\ref{fig:schematics_form}(c)), on which the gradient-based searches cannot find any effective descending directions to follow.

In this review, we will give a unified survey over the existing results on these different types of optimization landscapes. The remainder of this paper is organized as follows. Section~\ref{Sec:Problems} will formulate the landscape problems and how they arise in QOC and VQA applications. Section~\ref{Sec:Unlimited Systems} will analyze the landscape topology under unlimited resources, followed by Sec.~\ref{Sec:Limited Systems} which will introduce the change of topology when the resources are limited. In Sec.~\ref{Sec:BP}, the interesting Barren Plateau (BP) phenomena will be discussed as well as possible ways to avoid them. Finally, conclusions are drawn in Sec.~\ref{Sec:Conclusion}.

\section{Hybrid Quantum-Classical Optimization Problems}
\label{Sec:Problems}

As is schematically shown in Fig.~\ref{fig:schematics}, many quantum science and technology goals of fundamental and practical interest involve the directing quantum dynamics (e.g., molecules, spin ensembles, superconducting quantum devices) via iterative optimization of control fields, or variational optimization of parameterized quantum circuits. These problems commonly lead to the maximization or minimization of the expectation value of some objective observable $O$, which can be written as follows~\cite{Rabitz2006}
\begin{equation}
J(\theta) =\langle O\rangle = {\rm tr}\left[\rho(\theta)O\right],
\end{equation}
where $\rho(\theta)$ is the parameterized quantum state that encodes the solution to the problem. Here $\theta$ represents the practically manipulatable control variables for engineering the quantum system. Since the quantum state $\rho(\theta)$ is always prepared by some unitary transformation $U(\theta)$ on the system (assumed to be closed), we can alternatively write the objective function as
\begin{equation}\label{eq:landscape function}
J(\theta) = {\rm tr}\left[U(\theta)\rho_0 U^\dag(\theta)O\right],
\end{equation}
where $\rho_0$ is the initial state of the system. Here, the unitary transformation $U(\theta)$ represents the controlled dynamical propagator in QOC or the parameterized quantum circuit in VQA. In the following, we will show how these problems appear in quantum optimal control systems and variational quantum algorithms.

\subsection{Quantum optimal control systems}
The control of quantum systems exists in almost all quantum technologies. For example, shaped femtosecond laser pulses can be applied to manipulate chemical reactions by selectively breaking or forming chemical bonds~\cite{Assion1998,Daniel2003}. In quantum metrology, a quantum sensor can be optimally tuned to improve its sensitivity~\cite{Xu2021}. In quantum computing, such problems are prevalent because state initialization, gate operation and the suppression of noises can all be treated as control problems~\cite{Dong2010}.

Consider an ideal closed system where the general controlled quantum dynamics can be described by the following Schrodinger equation:
\begin{equation}\label{eq:quantum control system}
\dot{U}(t; \vec{u}) =-i\left[H_0+\sum_{k=1}^m u_k(t)H_k\right]U(t).
\end{equation}
Here, $U(t)$ is the system's unitary propagator and $H_0$ is the internal Hamiltonian. The control function $\vec{u}=\{u_1(t),\cdots,u_m(t)\}$ can be freely varied to manipulate the system via the respective control Hamiltonians $H_1,\cdots,H_k$. In many control problems, it is desired to find proper control functions such that
\begin{equation}\label{eq:QOC}
J(\vec{u}) = {\rm tr}\left[U(T;\vec{u})\rho_0 U^\dag(T;\vec{u})O\right]
\end{equation}
is extremized at some prescribed time $t=T$, where $\rho_0$ is the system's initial state. The observable may be a projector $O=|0\rangle\langle 0|$ onto the ground state, which appears in the initialization of a quantum information system. Let $\theta$ be the free parameters involved in the control function $\vec{u}$ (e.g., amplitudes and phases of piecewise-constant pulses), then the propagator $U(T;\vec{u})$ is implicitly parameterized as $U(T;\vec{u}(\theta))$ by $\theta$, and hence we can formulate (\ref{eq:QOC}) into the standard form (\ref{eq:landscape function}).

In quantum optimal control, the resources available for the optimizer depends on the number and range of variables in the control pulses, e.g., the time duration, the sampling rate, the power and the bandwidth, etc. The resources are also implicitly dependent on the coherence time of the system within which the propagator $U(T;\vec{u})$ in (\ref{eq:QOC}) is at least approximately unitary.

\subsection{Variational quantum algorithms}
Variational quantum algorithms can be deployed on any quantum system realization while exploiting quantum control resources. However, the use of NISQ computers may have a long term standing or shorter term period of utility before fault-tolerant quantum computers are available. For intermediate-scale or even large-scale quantum computers, parameterized quantum circuits (PQC) are broadly adopted as the computational ansatz with assistance of a classical optimizer, which are also referred to as quantum neural networks (QNN)~\cite{Benedetti2019}. Such computational models are expected to achieve a quantum advantage when there are sufficiently many qubits whose noises are sufficiently low and adequate control resources are available~\cite{Bishop2006}. PQC usually consists of layers of elementary single- or two-qubit gates, and part of them are tunable. For example, the single-qubit gates can be chosen as rotations around the $x$-axis on the Bloch sphere with the rotation angle being tunable. Let $\theta_1,\cdots,\theta_n$ be the parameters in these layers, and we have
\begin{equation}\label{}
U(\theta)=U_n(\theta_n)\cdots U_1(\theta_1).
\end{equation}

The architecture of the parameterized quantum circuits is determined by the qubit connectivity topology of the hardware device, and two-qubit gates are most convenient between directly coupled qubits. The selection of the quantum ansatz is sometimes inspired by the problem itself. For example, the quantum approximate optimization algorithm (QAOA)~\cite{Farhi2014,Farhi2019} uses alternating evolution of the initial and problem Hamiltonians $H_0$ and $H_P$, respectively, such that
\begin{equation}\label{}
U(\theta)=e^{-it_nH_P}e^{-i\tau_nH_P}\cdots e^{-it_1H_P}e^{-i\tau_1H_0},
\end{equation}
which is actually a quantum control system under bang-bang controls. The parameters $\theta$ consist of the evolution times $t_1,\tau_1,\cdots,t_n,\tau_n$. In practice, the problem-inspired ansatz may need to be transformed to hardware-inspired ones by decomposition and Trotterization.

The objective observable chosen for VQA depends on the specific applications. For a variational quantum eigensolver~\cite{Cao2019} or QAOA, the observable may be chosen as a non-local Hamiltonian encoding the problem that involves many qubit-qubit interactions. For machine learning tasks (e.g., classification), the observable can be defined locally on a few qubits whose states indicate the candidate output~\cite{Chen2021}. Later we will see that the choice of observables affects the landscape geometry.

Similar to QOC applications, the resources available for VQA are correlated with the tunable parameters involved in the quantum circuit, which is jointly determined by its width {(i.e., the number of qubits)} and number of layers. Due to the noise in NISQ devices, only shallow circuits can be properly utilized by the objective function (\ref{eq:landscape function}), otherwise any useful information will be buried in the noise.

\section{The Trap-free Landscape with Abundant Resources}
\label{Sec:Unlimited Systems}
In this section, we will show that the landscape is almost always trap-free under appropriate conditions and when the control resources are abundant. The analysis has been applied to explain the large number of successes in QOC experiments and simulations. The same method can be naturally generalized to small-scale quantum circuits by which any unitary can be achieved~\cite{Magann2021b}.

\subsection{Basic assumptions}
Suppose that the resources are abundant in the sense that any unitary $U$ can be realized by some properly chosen parameter $\theta\in\mathcal{X}$, i.e., the mapping $U(\theta)$ from $\mathcal{X}$ to the unitary group $\mathcal{U}(N)$ is surjective. This means that $M\geq N^2$ for physically suitable controls, and under many practical circumstances we actually have $M\gg N^2$. In control systems, this implies that the system is fully controllable over $\mathcal{U}(N)$, namely any unitary matrix can be produced by some control fields.

The presence of surjectivity makes it possible to transfer the landscape analysis to the following kinematic landscape
\begin{equation}\label{}
J(U)={\rm tr}[U\rho_0 U^\dag O],
\end{equation}
which is defined on the image of $\mathcal{X}$ that fills up $\mathcal{U}(N)$ under the resource-abundance assumption. This kinematic landscape is relatively easy to analyze because it is only quadratically dependent on $U$, while $J(\theta)$ may involve very complicated nonlinearities. The connection between the two landscapes can be understood from the chain rule for the $\alpha$-th control $\theta_\alpha$:
\begin{equation}\label{eq:chain1}
\frac{\partial J}{\partial \theta_\alpha} =  \sum_{i,j=1}^N\left[ \frac{\partial J}{\partial U_{ij}}\frac{\partial U_{ij}}{\partial \theta_\alpha}\right].
\end{equation}
In addition, at kinematic critical points where $\frac{\partial J}{\partial U(\theta)}=0$, the second-order derivatives are connected by
\begin{equation}\label{eq:chain2}
\frac{\partial^2 J}{\partial \theta_\alpha\partial\theta_\beta} =   \sum_{i,j=1}^N \sum_{k,l=1}^N\left[\frac{\partial U_{ij}}{\partial \theta_\alpha} \frac{\partial^2 J}{\partial U_{ij}\partial U_{kl}}\frac{\partial U_{kl}}{\partial \theta_\beta}\right],
\end{equation}

Clearly, the vanishing of the kinematic gradient $\frac{\partial J}{\partial U(\theta)}$ must lead to $\frac{\partial J}{\partial \theta} \equiv 0$, which implies that $\theta$ must be critical if the corresponding $U(\theta)$ is a critical point of $J(U)$. However, not all critical controls $\theta$ come from kinematic points, because the kinematic gradient can be nonzero when the Frechet derivative $\frac{\partial U(\theta)}{\partial \theta}$ is rank-deficient. Therefore, we conclude that the landscape topology is equivalent with the kinematic one under the following assumptions.

(1) The mapping from $\theta$ to $U$ is globally surjective, i.e., any unitary $U$ can be realized by some admissible $\theta$~\cite{HUANG1983}.

(2) The mapping from $\theta$ to $U$ is everywhere locally surjective, i.e., the Jacobian $\frac{\partial U}{\partial \theta}$ is full rank for all admissible $\theta$~\cite{Wu2012a}.

The assumptions also guarantee that, up to second order, any locally maximal (locally minimal) or saddle critical point must correspond to a kinematic critical point of the same type, because Eq.~(\ref{eq:chain2}) defines a congruent transformation that preserves the sign of non-zero Hessian eigenvalues, as long as the Jacobian mapping is full rank at the critical point. Theoretically, the second-order analysis is incomplete because higher-order variations can matter along directions associated with zero Hessian eigenvalues, implying that additional critical points that are unseen in the kinematic picture may exist due to the violation of local regularity. Nevertheless, both theoretical analysis and empirical studies indicate that such critical points are rare and have negligible influence on the search for globally optimal controls~\cite{Wu2012a,Riviello2014}.

\subsection{The critical topology of the landscape}
The above analysis shows that, as long as the classical optimizer has sufficiently abundant resources, generic landscape features can be extracted from the kinematic landscape. It is easy to prove that the condition for a unitary transformation $U$ to be a kinematic critical point is
\begin{equation}\label{}
[U\rho_0U^\dag,O]=0.
\end{equation}
At the critical point, the Hessian form is
\begin{equation}\label{}
\mathcal{H}(A)={\rm tr}(AU\rho_0U^\dag O-A^2U\rho_0U^\dag),
\end{equation}
which can be obtained from Taylor expanding $J(U)$ in the neighborhood of $U$ parameterized by $Ue^{iA}$ with $A$ being Hermitian and of small norm.
These conditions provide the basis for extracting all possible kinematic critical points and the curvature near them via Hessian analysis.

It is revealed that the kinematic landscape possesses a number of critical submanifolds among which only one is locally maximal (or minimal)~\cite{Rabitz2006,Rabitz2005}. The absence of other locally suboptimal extrema indicates that the gradient-based optimization of $\theta$ starting from an arbitrary point should almost always reach the top (or bottom) of the landscape without being trapped at lower (or upper) suboptimal values. There is no definite conclusion on the connectedness of the dynamical submanifolds, but most simulations appear to support that the maximal (minimal) submanifold is very likely connected, which can be numerically detected by the level-set exploration technique~\cite{Beltrani2007,Beltrani2011a} using homotopy algorithms\cite{Rothman2005,Dominy2008}.

In addition to the unique maximal and minimal submanifolds, there are usually multiple saddle submanifolds of $\mathcal{U}(N)$. Generically, there are fewer saddle submanifolds when $\rho_0$ or $O$ is highly degenerate, and these submanifolds tend to be high dimensional. Typically, when $\rho_0$ and $O$ are both fully non-degenerate, there is a total of $N!$ critical submanifolds, among which $N!-2$ are saddle submanifolds~\cite{Rabitz2006,Wu2008a}. However, when $\rho_0$ is a pure state, there are at most $N$ critical submanifolds, among which $N-2$ are saddles. For general cases, the contingency table technique~\cite{Wu2008a} was proposed to explicitly enumerate the critical submanifolds. Although the resulting combinatorial problem has no general analytic solutions, some special cases can be approximately estimated to get a good understanding of the distribution of critical manifolds~\cite{Wu2008a,Wu2008}.

\subsection{Fundamental bounds}
The bounds on the value of the objective function and the curvature are fundamental for the geometric understanding of the optimization landscape. They scale with the dimensionality of the quantum ansatz as well as the available resources of the classical optimizer. Here, we introduce some existing results on such bounds for landscapes with unlimited control resources.

It is clear that the value of the objective function is ultimately bounded by the maximal and minimal eigenvalues of the observable $O$. They are achievable when $\rho_0$ is a pure state and any unitary $U$ can be realized, but that may not always be the case when $\rho_0$ is mixed. Nevertheless, the presence of an ancillary system as a quantum controller (e.g., an engineerable environment for a control system or ancilla qubits for PQC) may purify the system's mixed state and thus broaden the achievable bounds of the landscape. Typically, when the ancillary system is initially at a state of thermal equilibrium, and the temperature decreases from infinite to zero, it was proven that the bounds limited by the purity of $\rho_0$ can be surpassed when the temperature is below some threshold value determined by the minimal energy gap of the Hamiltonian of the environment $H_E$, and the ultimate bound can be approached when the temperature goes to absolute zero. The threshold temperature can be taken as a witness index of the quantum effect of the environment, and the minimal energy gap of $H_E$ can be treated as the ``bandwidth" of the quantum controller that quantifies the ability of performance improvements~\cite{Wu2015aa}.

For the optimal control of quantum systems, the norm of the gradient is upper bounded by the product of the operator norms of $O$ and the control Hamiltonians, implying that the gradient-based search will never explode~\cite{Ho2006}. In other words, the landscape has limited slope. Moreover, when the system's dimension increases, any gradient component may shrink into an extremely narrow distribution centered at zero, leading to the so-called barren-plateau landscape~\cite{McClean2018}. On the plateau, it is exponentially expensive to precisely evaluate the gradient by sampling. Also, moving along the gradient will be very slow unless exponentially many control resources are available. More details of barren plateaus will be discussed in Sec.~\ref{Sec:BP}.

The flatness of the landscape can be also observed from the curvature associated with the Hessian form~\cite{Beltrani2011}, as gradient-based algorithms converge faster along Hessian eigenvectors associated with large Hessian eigenvalues. However, large Hessian eigenvalues also imply that the control is less robust to noise varying along the associated eigenvectors. Thus, a trade-off needs to be made between the convergence speed and the noise robustness~\cite{Hocker2014}.

\section{The Rugged Landscape with Limited Resources}
\label{Sec:Limited Systems}
The above analysis indicates that false traps (see Fig.~\ref{fig:schematics_form}(b)) will likely emerge when the classical optimizer does not have sufficiently many parameters to generate arbitrary quantum unitaries. In this section, we discuss how the landscape is reshaped under insufficient optimization resources.

From the control system point of view, the emergence of false traps is ascribed to the loss of controllability and regularity (regularity means that the Jacobian $\partial U/\partial \theta$ is full rank). The controllability of quantum systems can be examined by the rank of the Lie algebra generated by the drift and control Hamiltonians via their nested commutators. The system is controllable when the generated Lie algebra is identical with the Lie algebra ${\bf u}(N)$ of $\mathcal{U}(N)$~\cite{HUANG1983}. When the system possesses certain dynamic symmetry, i.e., when the generated Lie algebra is a proper Lie subalgebra of ${\bf u}(N)$, the system will become uncontrollable even when the associated control field resources are unlimited.

\begin{figure}
	\centering
	\includegraphics[width=1\columnwidth]{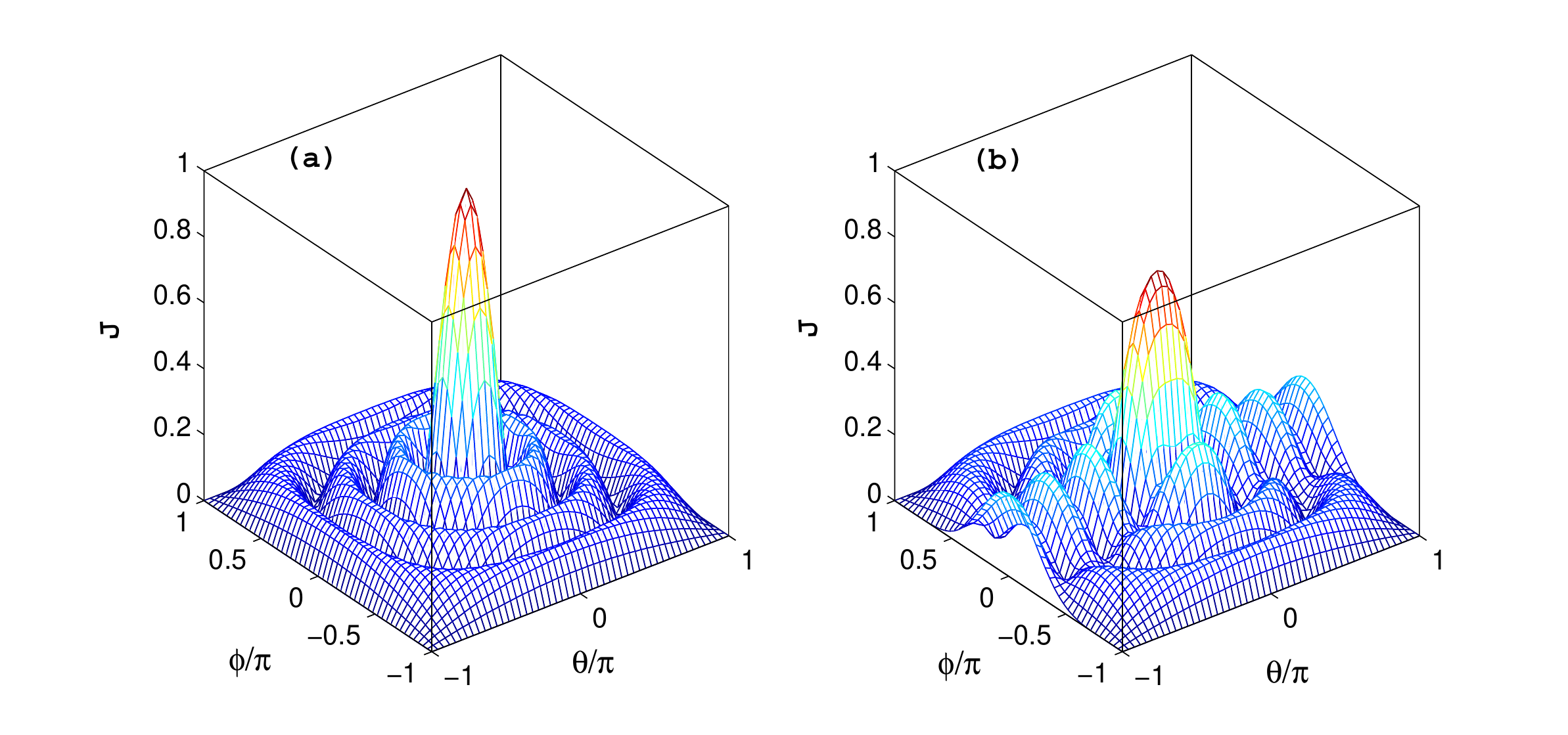}
	\caption{Local traps in the landscape induced by ${\bf SU}(2)$ dynamical symmetry reproduced with permission from Ref.~\cite{Wu2011}: (a) when the target gate is inside the ${\bf SU}(2)$ subgroup; (b) when the target gate is outside the ${\bf SU}(2)$ subgroup.}
	\label{fig3}
\end{figure}

In the literature, several cases of dynamical symmetry have been proven to introduce no traps for gate control landscapes, defined as $J(U)=Re{\rm tr}(W^\dag U)$ with $W$ being the target gate, including the set of symmetric unitary transformations, the set of symplectic dual transformations~\cite{Hsieh2010}, and the set of symplectic transformations in continuous-variable quantum computing systems~\cite{Wu2010}. However, false traps may appear when the generated Lie algebra is relatively small. In Ref.~\cite{Wu2011}, it is explicitly shown that $N/2$ physically nontrivial traps exist under $\mathcal{SU}(2)$ dynamic symmetry when the target gate $W$ is reachable (see Fig.~\ref{fig3}(a)), and the landscape usually becomes more rugged when $W$ is not reachable (see Fig.~\ref{fig3}b).

The loss of controllability and regularity are more commonly caused by physical constraints on the control field, even if the controllability Lie algebra is full rank. The constraints may be in the form of bounded power, finite bandwidth, etc., or the length of pulse duration limited by the coherence time. A simple way to systematically explore the landscape with constrained control resources is by restricting kinematic controls (e.g., entries of $U(T)$) that can be mapped to corresponding dynamic controls via a topology-preserving transformation. Suboptimal dynamic controls are identified as isolated points on the landscape, and they are shown to have rich and complex features~\cite{Donovan2013,Donovan2014a,Donovan2015}. Numerical simulations~\cite{Riviello2015} show that the search for a globally optimal solution may be prevented when constraints are above certain thresholds, and thus careful choice of relevant control parameters helps to eliminate such traps and facilitate successful optimization.

In variational quantum algorithms, uncontrollable quantum dynamics generally corresponds to under-parameterized quantum circuits such that reachability is in deficit. Hence, one can expect that the optimization landscape will likely be rugged as well. {Actually, the under-parameterized quantum circuits commonly have spurious local optima (i.e., traps) and thus the classical gradient-based optimizers may fail to achieve a globally optimal solution. For example, in Ref.~\cite{You2021}, a class of simple QNNs is identified to be hard to train, and there exist datasets that induce many spurious local suboptima that are exponential in the number of control variables. In such circumstances, the optimizer and the relevant algorithmic hyperparameters (e.g., the learning rate) can be carefully choosed to escape traps~\cite{Wierichs2020}. The undesired local suboptima can be also avoided by connecting the quantum circuits with a classical feedforward neural network which is expected to modify the landscape itself~\cite{Rivera2021}.} In noisy quantum devices, the landscape suffers more heavily from local traps, because the noise (e.g., non-unital Pauli noise) can break the symmetries in under-parameterized quantum circuits and lift the degeneracy of minima, making many of them false traps. Hence, novel optimization methods, e.g., the symmetry-based Minima Hopping (SYMH) optimizer, are required to mitigate the effect of noise and guide the search to more noise-resilient minima~\cite{Fontana2020}.

{In the literature, the influence of the number of quantum circuit parameters upon the landscape has been studied. It is noted that a few specific structured QNNs exhibit the over-parametrization phenomenon~\cite{Lee2021,Anschuetz2021,Larocca2021}, which is commonly observed in classical neural networks. The over-parametrization means that the QNN has more than a critical number of parameters which guarantees that the achievable rank of the quantum Fisher information matrix (QFIM) can be saturated at least at a point of the landscape. Numerical simulations show that the QFIM rank is saturated almost everywhere simultaneously. That is to say, the mapping from the parameters to the final state is surjective almost everywhere, leading to a favorable landscape. Thus, with increasing the number of the parameters, the QNN will experience a phase transition in trainability. For periodic-structured QNNs, it has been demonstrated that the critical threshold value of the parameter number is related to the dimension of the dynamical Lie Algebra (DLA) obtained from the QNN generators~\cite{Larocca2021}. For a quantum ansatz, deep layers needed to be over-parameterized may lead to barren plateaus. Therefore, the structure of quantum ansatz should be carefully designed to guarantee the scalability and trainability.  }

\section{The Barren-plateau Landscape with Scarce Resources}
\label{Sec:BP}
In practical applications (specially with VQAs), for an $n$ qubit system the number of tunable parameters is generally in $\mathcal{O}({\rm poly}(n))$ for the sake of computation efficiency, which is relatively scarce compared with the quantum system dimension when $n$ scales up. Under such circumstances, false traps may not be the major obstacle for optimization, because the presence of a barren plateau (BP) brings up greater challenges. In this section, we will discuss how BPs arise as well their origins and the ways to escape BPs.

\subsection{The effects of barren plateaus}

The BP phenomenon was first noticed in the study of quantum neural networks~\cite{McClean2018}. BP means that any gradient component has the zero mean value, i.e., ${\rm E}_{\theta} \left[\frac{\partial J}{\partial \theta_i}\right]=0$ for all $1\leq i\leq M$, over the parameter space $\mathcal{X}$ of the quantum ansatz, and its variance is exponentially bounded by
\begin{equation}
{\rm Var}_\theta\left(\frac{\partial J}{\partial \theta_i}\right)\leq e^{-\beta n}
\end{equation}
for some positive constant $\beta >0$. From Chebyshev's inequality, this further implies that the probability that the partial derivative differs from its zero mean at least by $\epsilon$ vanishes exponentially as
\begin{equation}
P\left(\left|\frac{\partial J}{\partial \theta_i}\right|\geq \epsilon\right)\leq \epsilon^{-2}e^{-\beta n}.
\end{equation}
This behavior indicates that, for sufficiently large qubit number $n$, the gradient is almost always vanishing at any randomly chosen $\theta$. In other words, almost all $\theta$ look like critical points, and when attempting to follow such gradients the search can hardly move. Consequently, a large number of iterations will be taken for training optimal parameters even in the best of circumstances when noise is very weak. Additionally, the exponential suppression of the gradient is often accompanied by exponentially narrowed minima~\cite{Arrasmith2021}, forcing the learning rate to be exponentially slow so as not to overstep the narrow gorge solutions. All these factors make the training extremely hard when the number of qubits is large.

Another side-effect of BPs is on the estimation of gradients (especially in VQA applications), which is often done via the parameter shift rule~\cite{Mitarai2018} as follows
\begin{equation}
\frac{\partial J(\theta)}{\partial \theta_j}=\frac{1}{2}\left[J\left(\theta_1,\cdots,\theta_j+\frac{\pi}{4},\cdots,\theta_M\right)-J\left(\theta_1,\cdots,\theta_j-\frac{\pi}{4},\cdots,\theta_M\right)\right].
\end{equation}
To guarantee a reliable gradient direction, an exponential number of repeated measurements on the objective function $J(\theta)$ is needed to overcome sampling noise. Otherwise, the optimization will perform no better than a random walk. This effect is unavoidable by simply changing the classical optimizer. In Ref.~\cite{Cerezo2021a}, it was shown that Newton-type algorithms make no difference because exponentially many measurements are still required to evaluate the Hessian matrix obtained by applying the parameter shift rule twice. In the presence of BPs, the landscape value will exhibit an exponential concentration about the mean~\cite{Arrasmith2021,Arrasmith2021b}, i.e., the variation ${\rm Var}_\theta[J(\theta)]$ exponentially decays with the qubit number $n$. Other gradient-free optimizers (e.g., Nelder-Mead, Powell, and COBYLA) cannot improve the efficiency of optimization, because decisions made in these algorithms are based on the comparison of the objective function values between different points~\cite{Arrasmith2021b}.

It should be noted that the exponential vanishing of the gradient components can be compensated for by adding exponentially many control parameters. In such a case, the gradients with a suitable norm could provide an effective direction to update the parameters $\theta$~\cite{Kim2021}. In addition, the landscape may not exhibit the narrow gorges or the value concentration phenomena, which generally occurs attendant on BPs when the control resources are scarce.

\subsection{The origins of barren plateaus}

There are multiple factors that form barren plateaus on the optimization landscapes of hybrid quantum-classical algorithms, including high expressibility of the quantum ansatz~\cite{McClean2018}, a non-locally defined objective function~\cite{Cerezo2021b,Sharma2020}, as well as excess entanglement~\cite{Patti2021,Marrero2021} and noises in the quantum ansatz~\cite{Wang2021,Franca2021}. {Besides the VQA landscapes, the quantum optimal control landscapes also suffer from the BPs with the scaling system dimension. As shown in Ref.~\cite{Arenz2020}, for the uniformly random target state generation problem, the control landscape is exponentially flat as a consequence of the concentration of measure.}

The first discovery of the BP phenomenon was in random and deep parameterized quantum circuits~\cite{McClean2018}, where BP was proven to exist when the unitary transformations produced by the PQC form a $2$-design, i.e., they are sufficiently random so that the average
\begin{equation}
\int_{\mathcal{X}} {\rm d}U(\theta) U(\theta)^{\otimes 2}\rho (U^\dagger(\theta))^{\otimes 2}=\int_{\mathcal{U}(N)}{\rm d}\mu (U)U^{\otimes 2}\rho (U^\dagger)^{\otimes 2}
\end{equation}
holds for any $\rho$, where $\mu (U)$ denotes the Haar distribution on the unitary group. In practice, a PQC can approximate a $2$-design when it is sufficiently deep, e.g., a hardware-efficient ansatz with depth $\mathcal{O}({\rm poly}(n))$~\cite{Renes2004,Harrow2009}. {Since the distance of the ansatz from being a 2-design measures its expressibility,} similar to the controllability of a quantum control system, the appearance of BPs indicates that a highly expressive quantum ansatz would exhibit a flat optimization landscape and the training gets harder~\cite{Holmes2022}.

\begin{figure}
	\centering
	\includegraphics[width=0.6\textwidth]{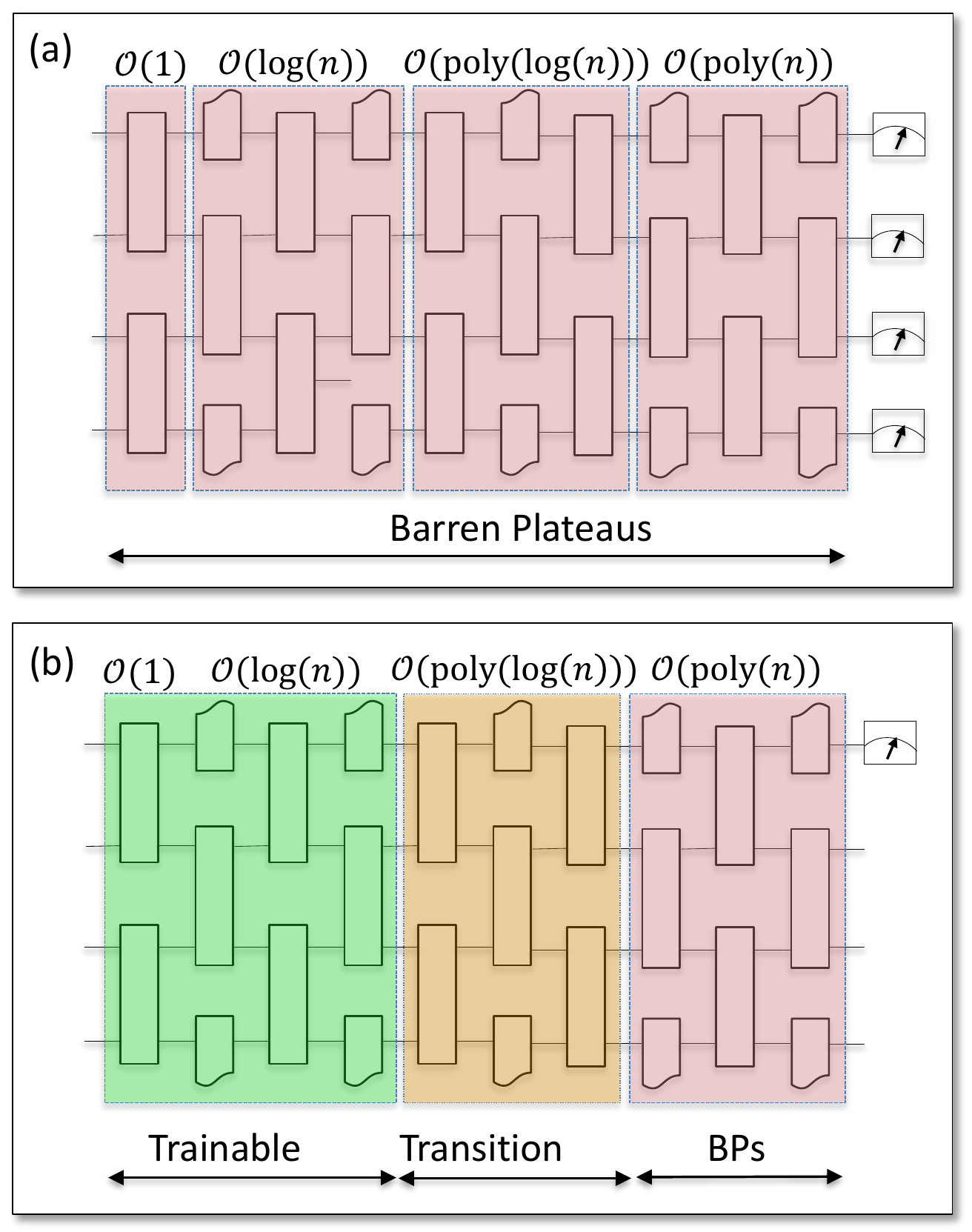}
	\caption{Trainability of the hardware-efficient ansatz in terms of the circuit depth for the global cost function (upper) and the local cost function (lower), respectively~\cite{Cerezo2021b}.}
	\label{fig:locality}
\end{figure}

The BP phenomenon can be observed in shallow quantum circuits with certain objective functions~\cite{Cerezo2021b,Uvarov2021}. As shown in Fig.~\ref{fig:locality}, global objective functions (i.e., when the observable $O$ non-trivially acts on all qubits) can lead to BPs for circuits at any depth as long as the layered hardware-efficient ansatz consists of blocks of local 2-designs. When the objective function is
locally defined (i.e., when $O$ acts only on a few qubits), the quantum ansatz is trainable when the circuit depth is at the level of $\mathcal{O}({\rm log}(n))$ because the gradient vanishes at worst polynomially, while BPs appear when the depth is polynomial in $n$. The dependence of the locality of the observable can be found for more general objective functions consisting of Pauli strings~\cite{Uvarov2021}, but the structure of the ansatz has even more subtle influence.

The connection between BPs and non-local objective functions can be understood from another perspective. Recently, it was found that BPs can be also induced by strong entanglement between qubits~\cite{Patti2021,Marrero2021}, the non-local correlation that is deemed as the most valuable resource for achieving advantageous quantum computing. For example, the QNN that satisfies a volume-law in the entanglement entropy is difficult to train because the gradient vanishes exponentially with the number of hidden qubits for any bounded observable operators. Like the above conflict with expressivity, this raises the trade-off between trainability and the entanglement resource that has to be made in practice.

On NISQ devices, the noise also induces BPs~\cite{Wang2021}, because the final quantum state processed by the ansatz exponentially converges to the maximally mixed state which leads to a flattening of the optimization landscape. Complexity analysis shows that, for the local Pauli noise that acts throughout the PQC, the gradient vanishes exponentially in the number of qubits $n$ when the depth of the ansatz is linear with $n$. However, it should be noted that the exponential decay is for the gradient itself, but not the variance of the gradient discussed above, and the decay is purely a decoherence effect that is independent of the parameter initialization strategy, the locality of the objective funciton, or the structure of ansatz.

\subsection{Algorithm design for mitigating barren-plateau effects}

Based on the above understanding of BPs, several strategies have been proposed to avoid or mitigate the effects of the BPs on optimization.

\begin{figure}
	\centering
	\includegraphics[width=0.6\textwidth]{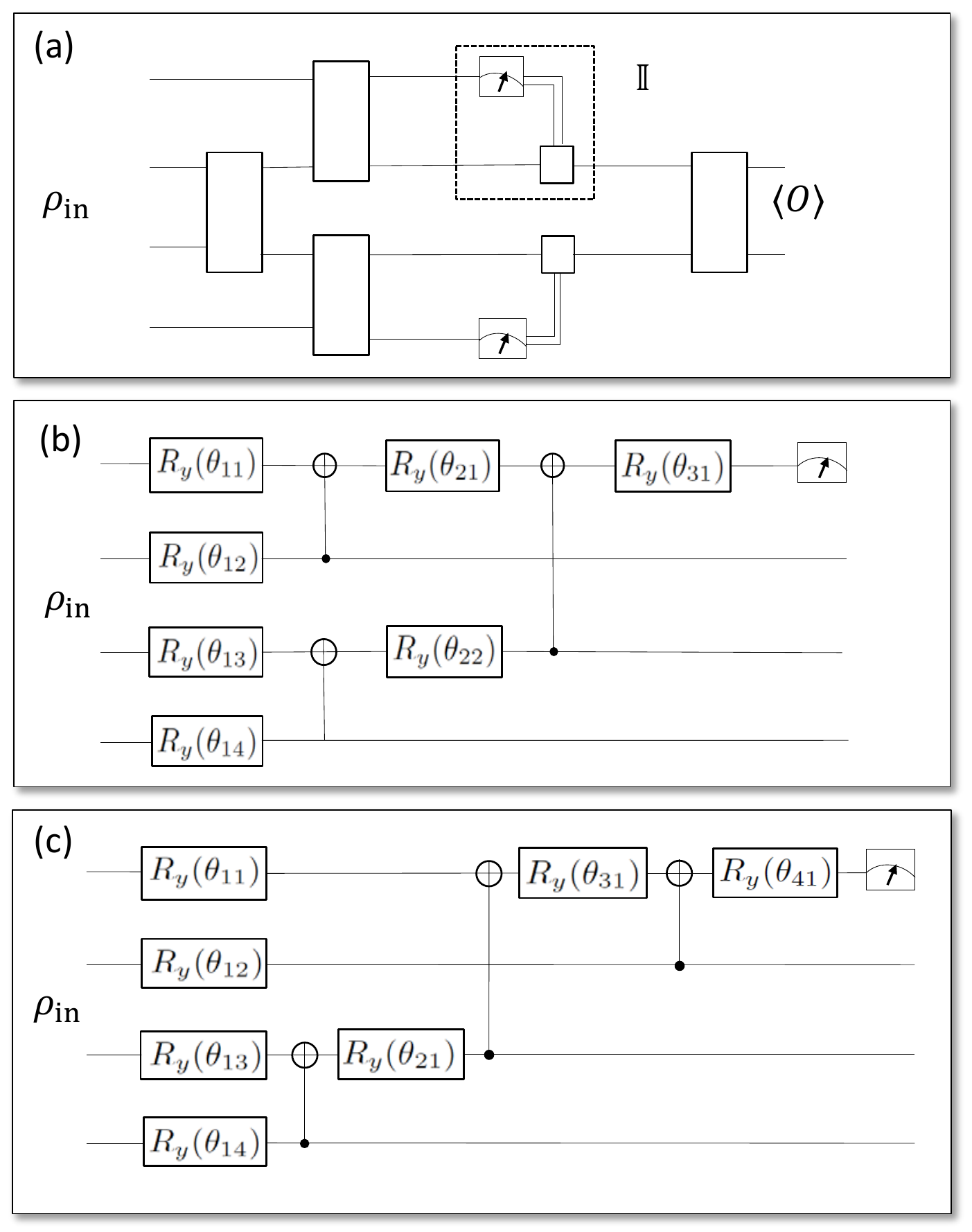}
	\caption{Quantum neural networks with special structure: (a) quantum convolutional neural network involving a sequence of convolutional and pooling layers~\cite{Pesah2021}; (b) quantum neural network with a tree tensor structure~\cite{Zhang2020}; (c) quantum neural network with a step controlled structure~\cite{Zhang2020}.}
	\label{fig:QNN}
\end{figure}

Well-designed PQC architectures can be resilient to the BP phenomenon. In Ref.~\cite{Nakaji2021}, it was shown that the expressibility and the trainability can coexist for a class of shallow alternating layered ansatzs. As shown in Fig.~\ref{fig:QNN}, high trainability can be achieved in QNNs with at most a polynomially decaying gradient~\cite{Pesah2021,Zhang2020}, such as the convolutional QNN and QNN with a tree tensor structure or with a step controlled structure. {Another special structure, namely system-agnostic ansatzs based on trainable Fourier coefficients of Hamiltonian system parameters, has been also shown to be mild or entirely absent from BPs ~\cite{Broers2021}. In addition, it is noted that for a periodic-structured problem-inspired ansatz the variance of the partial derivative is inversely proportional to the dimension of the DLA and thus the gradient scaling can be diagnosed by the degree of the system controllability~\cite{Larocca2021b}. This provides a better understanding and predication {for the presence or absence of BPs} in problem-inspired ansatzs such as QAOA and the Hamiltonian variational ansatz (HVA)~\cite{Larocca2021b,Wiersema2020}.}

In addition to the structure design, BPs can also be avoided by specially designed initialization schemes. For the example of QAOA, a high-quality parameter initialization can be obtained by solving a similar but smaller-size problem owing to the widespread parameter concentration phenomenon~\cite{Brandao2018,Rudolph2021}. For general PQCs, it is proposed that one can randomly select some of the initial parameter values and fit the remaining values such that the result is a fixed unitary matrix~\cite{Grant2019}. The main idea thereof is to restrict the randomness and circuit depth so that it cannot approach a $2$-design. Alternatively, one can reduce the dimensionality of the parameter space by using random PQC architectures containing correlated parameters, even when the objective function is defined by a global operator~\cite{Volkoff2021}. However, expressivity is sacrificed in the circumstance.

Similar to the training of classical deep neural networks, the layerwise training strategy has also been proposed to avoid the problem of BPs because each training stage addresses only a low-depth circuit~\cite{Skolik2021}.
In Ref.~\cite{Verdon2019}, the idea of pretraining is introduced, i.e., the parameters are pretrained by classical neural networks, which are then transferred to the quantum neural network and fine tuned. In this way, the total number of optimization iterations required to reach a given accuracy can be significantly improved.

Both the initialization and ansatz design strategies are useless for noise-induced BPs. In such a case, efficient error mitigation techniques~\cite{Temme2017,Endo2021} become an indispensable ingredient to improve trainability. As an example, a VQE combined with advanced error mitigation strategies was applied to accurately model the binding energy of hydrogen chains, which uses up to a dozen qubits on the Sycamore quantum processor~\cite{Arute2020}.

\section{Conclusion}
\label{Sec:Conclusion}
To conclude, we have reviewed existing studies on the optimization landscape of hybrid quantum-classical algorithms that consist of a quantum ansatz and a classical optimizer. In the greater view that ranges from small-scale quantum control systems to intermediate- and then large-scale quantum circuits, a morphological transition of the landscape is displayed when the optimizer turns from resource-abundant to resource-scarce with respect to the exponentially increasing size of the ansatz.

These results explain why the control of small-size quantum systems was so successful over the past decades, and how hard it will be to work with NISQ devices and algorithms. In particular, the established landscape for high-dimensional quantum systems shows that a compromise needs to be made between the expressivity (or the controllability in the context of control) of the quantum ansatz and the trainability (or the efficiency) of the optimizer. This conclusion implies that, with regard to practical applications, quantum advantages that are expected with NISQ devices may not be easy to find. A better understanding of the ansatz-optimizer trade-off between the quantum and classical players may provide clearer path towards maximally extracting the power of NISQ algorithms. Many open challenges are ahead to explore.

Further into the future, the hybrid quantum-classical algorithms may gradually evolve into full quantum-quantum algorithms where the optimizer is realized by a fault-tolerant programmable quantum computer. In this scenario, the optimizer may possess equivalently as many resources as the ansatz, and the underlying landscape may look very different. Some relevant investigations have been done from the view of convex optimization~\cite{Wu2019,Banchi2020}, but there are many more directions ahead to explore.

\section*{References}
\bibliography{landscape}

\begin{thebibliography}{10}
\expandafter\ifx\csname url\endcsname\relax
  \def\url#1{\texttt{#1}}\fi
\expandafter\ifx\csname urlprefix\endcsname\relax\def\urlprefix{URL }\fi
\expandafter\ifx\csname href\endcsname\relax
  \def\href#1#2{#2} \def\path#1{#1}\fi

\bibitem{Magann2021}
A.~B. Magann, M.~D. Grace, H.~A. Rabitz, M.~Sarovar,
  \href{https://link.aps.org/doi/10.1103/PhysRevResearch.3.023165}{Digital
  quantum simulation of molecular dynamics and control}, Phys. Rev. Research 3
  (2021) 023165.
\newblock \href {http://dx.doi.org/10.1103/PhysRevResearch.3.023165}
  {\path{doi:10.1103/PhysRevResearch.3.023165}}.
\newline\urlprefix\url{https://link.aps.org/doi/10.1103/PhysRevResearch.3.023165}

\bibitem{Degen2017}
C.~L. Degen, F.~Reinhard, P.~Cappellaro,
  \href{https://link.aps.org/doi/10.1103/RevModPhys.89.035002}{Quantum
  sensing}, Rev. Mod. Phys. 89 (2017) 035002.
\newblock \href {http://dx.doi.org/10.1103/RevModPhys.89.035002}
  {\path{doi:10.1103/RevModPhys.89.035002}}.
\newline\urlprefix\url{https://link.aps.org/doi/10.1103/RevModPhys.89.035002}

\bibitem{Poggiali2018}
F.~Poggiali, P.~Cappellaro, N.~Fabbri,
  \href{https://link.aps.org/doi/10.1103/PhysRevX.8.021059}{Optimal control for
  one-qubit quantum sensing}, Phys. Rev. X 8 (2018) 021059.
\newblock \href {http://dx.doi.org/10.1103/PhysRevX.8.021059}
  {\path{doi:10.1103/PhysRevX.8.021059}}.
\newline\urlprefix\url{https://link.aps.org/doi/10.1103/PhysRevX.8.021059}

\bibitem{Waks2002}
E.~Waks, K.~Inoue, C.~Santori, D.~Fattal, J.~Vuckovic, G.~S. Solomon,
  Y.~Yamamoto, \href{https://doi.org/10.1038/420762a}{Quantum cryptography with
  a photon turnstile}, Nature 420~(6917) (2002) 762--762.
\newblock \href {http://dx.doi.org/10.1038/420762a}
  {\path{doi:10.1038/420762a}}.
\newline\urlprefix\url{https://doi.org/10.1038/420762a}

\bibitem{Resch2019}
S.~Resch, U.~R. Karpuzcu, Quantum computing: An overview across the system
  stack (2019).
\newblock \href {http://arxiv.org/abs/1905.07240} {\path{arXiv:1905.07240}}.

\bibitem{Brif2010}
C.~Brif, R.~Chakrabarti, H.~Rabitz, Control of quantum phenomena: past, present
  and future, New Journal of Physics 12 (2010) 075008.
\newblock \href {http://dx.doi.org/10.1088/1367-2630/12/7/075008}
  {\path{doi:10.1088/1367-2630/12/7/075008}}.

\bibitem{Glaser2015}
S.~J. Glaser, U.~Boscain, T.~Calarco, C.~P. Koch, W.~Kockenberger, R.~Kosloff,
  I.~Kuprov, B.~Luy, S.~Schirmer, T.~Schulte-Herbruggen, D.~Sugny, F.~K.
  Wilhelm, Training schrodinger's cat: quantum optimal control strategic report
  on current status, visions and goals for research in europe, European
  Physical Journal D 69~(12) (2015) 279.
\newblock \href {http://dx.doi.org/10.1140/epjd/e2015-60464-1}
  {\path{doi:10.1140/epjd/e2015-60464-1}}.

\bibitem{Cerezo2021}
M.~Cerezo, A.~Arrasmith, R.~Babbush, S.~C. Benjamin, S.~Endo, K.~Fujii, J.~R.
  McClean, K.~Mitarai, X.~Yuan, L.~Cincio, P.~J. Coles, Variational quantum
  algorithms, Nature Reviews Physics 3~(9) (2021) 625--644.
\newblock \href {http://dx.doi.org/10.1038/s42254-021-00348-9}
  {\path{doi:10.1038/s42254-021-00348-9}}.

\bibitem{Preskill2018}
J.~Preskill, Quantum {C}omputing in the {NISQ} era and beyond, {Quantum} 2
  (2018) 79.

\bibitem{Lubasch2020}
M.~Lubasch, J.~Joo, P.~Moinier, M.~Kiffner, D.~Jaksch,
  \href{https://link.aps.org/doi/10.1103/PhysRevA.101.010301}{Variational
  quantum algorithms for nonlinear problems}, Phys. Rev. A 101 (2020) 010301.
\newblock \href {http://dx.doi.org/10.1103/PhysRevA.101.010301}
  {\path{doi:10.1103/PhysRevA.101.010301}}.
\newline\urlprefix\url{https://link.aps.org/doi/10.1103/PhysRevA.101.010301}

\bibitem{Bharti2021}
K.~Bharti, A.~Cervera-Lierta, T.~H. Kyaw, T.~Haug, S.~Alperin-Lea, A.~Anand,
  M.~Degroote, H.~Heimonen, J.~S. Kottmann, T.~Menke, W.-K. Mok, S.~Sim, L.-C.
  Kwek, A.~Aspuru-Guzik, Noisy intermediate-scale quantum (nisq) algorithms
  (2021).
\newblock \href {http://arxiv.org/abs/2101.08448} {\path{arXiv:2101.08448}}.

\bibitem{Benedetti2021}
M.~Benedetti, M.~Fiorentini, M.~Lubasch,
  \href{https://link.aps.org/doi/10.1103/PhysRevResearch.3.033083}{Hardware-efficient
  variational quantum algorithms for time evolution}, Phys. Rev. Research 3
  (2021) 033083.
\newblock \href {http://dx.doi.org/10.1103/PhysRevResearch.3.033083}
  {\path{doi:10.1103/PhysRevResearch.3.033083}}.
\newline\urlprefix\url{https://link.aps.org/doi/10.1103/PhysRevResearch.3.033083}

\bibitem{Plekhanov2022}
K.~Plekhanov, M.~Rosenkranz, M.~Fiorentini, M.~Lubasch,
  \href{https://doi.org/10.22331%2Fq-2022-03-17-670}{Variational quantum
  amplitude estimation}, Quantum 6 (2022) 670.
\newblock \href {http://dx.doi.org/10.22331/q-2022-03-17-670}
  {\path{doi:10.22331/q-2022-03-17-670}}.
\newline\urlprefix\url{https://doi.org/10.22331%2Fq-2022-03-17-670}

\bibitem{Amaro2022}
D.~Amaro, C.~Modica, M.~Rosenkranz, M.~Fiorentini, M.~Benedetti, M.~Lubasch,
  \href{https://doi.org/10.1088%2F2058-9565%2Fac3e54}{Filtering variational
  quantum algorithms for combinatorial optimization}, Quantum Science and
  Technology 7~(1) (2022) 015021.
\newblock \href {http://dx.doi.org/10.1088/2058-9565/ac3e54}
  {\path{doi:10.1088/2058-9565/ac3e54}}.
\newline\urlprefix\url{https://doi.org/10.1088%2F2058-9565%2Fac3e54}

\bibitem{Dou2022}
T.~Dou, G.~Zhang, W.~Cui, \href{https://arxiv.org/abs/2201.01246}{Efficient
  quantum feature extraction for cnn-based learning} (2022).
\newblock \href {http://dx.doi.org/10.48550/ARXIV.2201.01246}
  {\path{doi:10.48550/ARXIV.2201.01246}}.
\newline\urlprefix\url{https://arxiv.org/abs/2201.01246}

\bibitem{Judson1992}
R.~S. Judson, H.~Rabitz,
  \href{https://link.aps.org/doi/10.1103/PhysRevLett.68.1500}{Teaching lasers
  to control molecules}, Phys. Rev. Lett. 68 (1992) 1500--1503.
\newblock \href {http://dx.doi.org/10.1103/PhysRevLett.68.1500}
  {\path{doi:10.1103/PhysRevLett.68.1500}}.
\newline\urlprefix\url{https://link.aps.org/doi/10.1103/PhysRevLett.68.1500}

\bibitem{Rabitz2004}
H.~A. Rabitz, M.~M. Hsieh, C.~M. Rosenthal, Quantum optimally controlled
  transition landscapes, Science 303~(5666) (2004) 1998--2001.
\newblock \href {http://dx.doi.org/10.1126/science.1093649}
  {\path{doi:10.1126/science.1093649}}.

\bibitem{Riviello2017}
G.~Riviello, R.~B. Wu, Q.~Y. Sun, H.~Rabitz, Searching for an optimal control
  in the presence of saddles on the quantum-mechanical observable landscape,
  Physical Review A 95~(6) (2017) 063418.
\newblock \href {http://dx.doi.org/10.1103/PhysRevA.95.063418}
  {\path{doi:10.1103/PhysRevA.95.063418}}.

\bibitem{McClean2018}
J.~R. McClean, S.~Boixo, V.~N. Smelyanskiy, R.~Babbush, H.~Neven, Barren
  plateaus in quantum neural network training landscapes, Nature Communications
  9 (2018) 4812.
\newblock \href {http://dx.doi.org/10.1038/s41467-018-07090-4}
  {\path{doi:10.1038/s41467-018-07090-4}}.

\bibitem{Rabitz2006}
H.~Rabitz, M.~Hsieh, C.~Rosenthal, Optimal control landscapes for quantum
  observables, Journal of Chemical Physics 124~(20) (2006) 204107.
\newblock \href {http://dx.doi.org/10.1063/1.2198837}
  {\path{doi:10.1063/1.2198837}}.

\bibitem{Assion1998}
A.~Assion, T.~Baumert, M.~Bergt, T.~Brixner, B.~Kiefer, V.~Seyfried,
  M.~Strehle, G.~Gerber,
  \href{https://www.science.org/doi/abs/10.1126/science.282.5390.919}{Control
  of chemical reactions by feedback-optimized phase-shaped femtosecond laser
  pulses}, Science 282~(5390) (1998) 919--922.
\newblock \href
  {http://arxiv.org/abs/https://www.science.org/doi/pdf/10.1126/science.282.5390.919}
  {\path{arXiv:https://www.science.org/doi/pdf/10.1126/science.282.5390.919}},
  \href {http://dx.doi.org/10.1126/science.282.5390.919}
  {\path{doi:10.1126/science.282.5390.919}}.
\newline\urlprefix\url{https://www.science.org/doi/abs/10.1126/science.282.5390.919}

\bibitem{Daniel2003}
C.~Daniel, J.~Full, L.~Gonz\'{a}lez, C.~Lupulescu, J.~Manz, A.~Merli,
  \v{S}tefan Vajda, L.~W\''{o}ste,
  \href{https://www.science.org/doi/abs/10.1126/science.1078517}{Deciphering
  the reaction dynamics underlying optimal control laser fields}, Science
  299~(5606) (2003) 536--539.
\newblock \href
  {http://arxiv.org/abs/https://www.science.org/doi/pdf/10.1126/science.1078517}
  {\path{arXiv:https://www.science.org/doi/pdf/10.1126/science.1078517}}, \href
  {http://dx.doi.org/10.1126/science.1078517}
  {\path{doi:10.1126/science.1078517}}.
\newline\urlprefix\url{https://www.science.org/doi/abs/10.1126/science.1078517}

\bibitem{Xu2021}
H.~Xu, L.~Wang, H.~Yuan, X.~Wang,
  \href{http://dx.doi.org/10.1103/PhysRevA.103.042615}{Generalizable control
  for multiparameter quantum metrology}, Physical Review A 103~(4).
\newblock \href {http://dx.doi.org/10.1103/physreva.103.042615}
  {\path{doi:10.1103/physreva.103.042615}}.
\newline\urlprefix\url{http://dx.doi.org/10.1103/PhysRevA.103.042615}

\bibitem{Dong2010}
D.~Dong, I.~Petersen,
  \href{http://dx.doi.org/10.1049/iet-cta.2009.0508}{Quantum control theory and
  applications: a survey}, IET Control Theory \& Applications 4~(12) (2010)
  2651--2671.
\newblock \href {http://dx.doi.org/10.1049/iet-cta.2009.0508}
  {\path{doi:10.1049/iet-cta.2009.0508}}.
\newline\urlprefix\url{http://dx.doi.org/10.1049/iet-cta.2009.0508}

\bibitem{Benedetti2019}
M.~Benedetti, E.~Lloyd, S.~Sack, M.~Fiorentini, Parameterized quantum circuits
  as machine learning models, arXiv preprint~(1906.07682).
\newblock \href {http://arxiv.org/abs/arXiv:1906.07682}
  {\path{arXiv:arXiv:1906.07682}}.

\bibitem{Bishop2006}
C.~M. Bishop, Pattern Recognition and Machine Learning (Information Science and
  Statistics), Springer-Verlag New York, Inc., 2006.

\bibitem{Farhi2014}
E.~Farhi, J.~Goldstone, S.~Gutmann, A quantum approximate optimization
  algorithm (2014).
\newblock \href {http://arxiv.org/abs/1411.4028} {\path{arXiv:1411.4028}}.

\bibitem{Farhi2019}
E.~Farhi, A.~W. Harrow, Quantum supremacy through the quantum approximate
  optimization algorithm (2019).
\newblock \href {http://arxiv.org/abs/1602.07674} {\path{arXiv:1602.07674}}.

\bibitem{Cao2019}
Y.~Cao, J.~Romero, J.~P. Olson, M.~Degroote, P.~D. Johnson, M.~Kieferov\'{a},
  I.~D. Kivlichan, T.~Menke, B.~Peropadre, N.~P.~D. Sawaya, S.~Sim, L.~Veis,
  A.~Aspuru-Guzik, Quantum chemistry in the age of quantum computing, Chem.
  Rev. 119~(19) (2019) 10856--10915.
\newblock \href {http://dx.doi.org/10.1021/acs.chemrev.8b00803}
  {\path{doi:10.1021/acs.chemrev.8b00803}}.

\bibitem{Chen2021}
S.~Y.-C. Chen, C.-M. Huang, C.-W. Hsing, Y.-J. Kao,
  \href{http://dx.doi.org/10.1088/2632-2153/ac104d}{An end-to-end trainable
  hybrid classical-quantum classifier}, Machine Learning: Science and
  Technology 2~(4) (2021) 045021.
\newblock \href {http://dx.doi.org/10.1088/2632-2153/ac104d}
  {\path{doi:10.1088/2632-2153/ac104d}}.
\newline\urlprefix\url{http://dx.doi.org/10.1088/2632-2153/ac104d}

\bibitem{Magann2021b}
A.~B. Magann, C.~Arenz, M.~D. Grace, T.-S. Ho, R.~L. Kosut, J.~R. McClean,
  H.~A. Rabitz, M.~Sarovar,
  \href{https://doi.org/10.1103%2Fprxquantum.2.010101}{From pulses to circuits
  and back again: A quantum optimal control perspective on variational quantum
  algorithms}, {PRX} Quantum 2~(1).
\newblock \href {http://dx.doi.org/10.1103/prxquantum.2.010101}
  {\path{doi:10.1103/prxquantum.2.010101}}.
\newline\urlprefix\url{https://doi.org/10.1103%2Fprxquantum.2.010101}

\bibitem{HUANG1983}
G.~M. Huang, T.~J. Tarn, J.~W. Clark, On the controllability of
  quantum-mechanical systems, Journal of Mathematical Physics 24~(11) (1983)
  2608--2618.
\newblock \href {http://dx.doi.org/10.1063/1.525634}
  {\path{doi:10.1063/1.525634}}.

\bibitem{Wu2012a}
R.~B. Wu, R.~X. Long, J.~Dominy, T.~S. Ho, H.~Rabitz, Singularities of quantum
  control landscapes, Physical Review A 86~(1) (2012) 013405.
\newblock \href {http://dx.doi.org/10.1103/PhysRevA.86.013405}
  {\path{doi:10.1103/PhysRevA.86.013405}}.

\bibitem{Riviello2014}
G.~Riviello, C.~Brif, R.~X. Long, R.~B. Wu, K.~M. Tibbetts, T.~S. Ho,
  H.~Rabitz, Searching for quantum optimal control fields in the presence of
  singular critical points, Physical Review A 90~(1) (2014) 013404.
\newblock \href {http://dx.doi.org/10.1103/PhysRevA.90.013404}
  {\path{doi:10.1103/PhysRevA.90.013404}}.

\bibitem{Rabitz2005}
H.~Rabitz, M.~Hsieh, C.~Rosenthal, Landscape for optimal control of
  quantum-mechanical unitary transformations, Physical Review A 72~(5) (2005)
  052337.
\newblock \href {http://dx.doi.org/10.1103/PhysRevA.72.052337}
  {\path{doi:10.1103/PhysRevA.72.052337}}.

\bibitem{Beltrani2007}
V.~Beltrani, J.~Dominy, T.-S. Ho, H.~Rabitz, Photonic reagent control of
  dynamically homologous quantum systems, Journal of Chemical Physics 126~(9)
  (2007) 094105.
\newblock \href {http://dx.doi.org/10.1063/1.2434177}
  {\path{doi:10.1063/1.2434177}}.

\bibitem{Beltrani2011a}
V.~Beltrani, J.~Dominy, T.~S. Ho, H.~Rabitz, Exploring the top and bottom of
  the quantum control landscape, Journal of Chemical Physics 134~(19) (2011)
  194106.
\newblock \href {http://dx.doi.org/10.1063/1.3589404}
  {\path{doi:10.1063/1.3589404}}.

\bibitem{Rothman2005}
A.~Rothman, T.~S. Ho, H.~Rabitz, Quantum observable homotopy tracking control,
  Journal of Chemical Physics 123~(13) (2005) 134104.
\newblock \href {http://dx.doi.org/10.1063/1.2042456}
  {\path{doi:10.1063/1.2042456}}.

\bibitem{Dominy2008}
J.~Dominy, H.~Rabitz, Exploring families of quantum controls for generating
  unitary transformations, Journal of Physics A-mathematical and Theoretical
  41~(20) (2008) 205305.
\newblock \href {http://dx.doi.org/10.1088/1751-8113/41/20/205305}
  {\path{doi:10.1088/1751-8113/41/20/205305}}.

\bibitem{Wu2008a}
R.~Wu, H.~Rabitz, M.~Hsieh, Characterization of the critical submanifolds in
  quantum ensemble control landscapes, Journal of Physics A-mathematical and
  Theoretical 41~(1) (2008) 015006.
\newblock \href {http://dx.doi.org/10.1088/1751-8113/41/1/015006}
  {\path{doi:10.1088/1751-8113/41/1/015006}}.

\bibitem{Wu2008}
R.~Wu, A.~Pechen, H.~Rabitz, M.~Hsieh, B.~Tsou, Control landscapes for
  observable preparation with open quantum systems, Journal of Mathematical
  Physics 49~(2) (2008) 022108.
\newblock \href {http://dx.doi.org/10.1063/1.2883738}
  {\path{doi:10.1063/1.2883738}}.

\bibitem{Wu2015aa}
R.-B. Wu, C.~Brif, M.~R. James, H.~Rabitz,
  \href{https://link.aps.org/doi/10.1103/PhysRevA.91.042327}{Limits of optimal
  control yields achievable with quantum controllers}, Phys. Rev. A 91 (2015)
  042327.
\newblock \href {http://dx.doi.org/10.1103/PhysRevA.91.042327}
  {\path{doi:10.1103/PhysRevA.91.042327}}.
\newline\urlprefix\url{https://link.aps.org/doi/10.1103/PhysRevA.91.042327}

\bibitem{Ho2006}
T.-S. Ho, H.~Rabitz, Why do effective quantum controls appear easy to find?,
  Journal of Photochemistry and Photobiology A-chemistry 180~(3) (2006)
  226--240.
\newblock \href {http://dx.doi.org/10.1016/j.jphotochem.2006.03.038}
  {\path{doi:10.1016/j.jphotochem.2006.03.038}}.

\bibitem{Beltrani2011}
V.~Beltrani, J.~Dominy, T.~S. Ho, H.~Rabitz, Bounds on the curvature at the top
  and bottom of the transition probability landscape, Journal of Physics
  B-atomic Molecular and Optical Physics 44~(15) (2011) 154009.
\newblock \href {http://dx.doi.org/10.1088/0953-4075/44/15/154009}
  {\path{doi:10.1088/0953-4075/44/15/154009}}.

\bibitem{Hocker2014}
D.~Hocker, C.~Brif, M.~D. Grace, A.~Donovan, T.~S. Ho, K.~M. Tibbetts, R.~B.
  Wu, H.~Rabitz, Characterization of control noise effects in optimal quantum
  unitary dynamics, Physical Review A 90~(6) (2014) 062309.
\newblock \href {http://dx.doi.org/10.1103/PhysRevA.90.062309}
  {\path{doi:10.1103/PhysRevA.90.062309}}.

\bibitem{Hsieh2010}
M.~Hsieh, R.~B. Wu, H.~Rabitz, D.~Lidar, Optimal control landscape for the
  generation of unitary transformations with constrained dynamics, Physical
  Review A 81~(6) (2010) 062352.
\newblock \href {http://dx.doi.org/10.1103/PhysRevA.81.062352}
  {\path{doi:10.1103/PhysRevA.81.062352}}.

\bibitem{Wu2010}
R.~B. Wu, R.~Chakrabarti, H.~Rabitz, Critical landscape topology for
  optimization on the symplectic group, Journal of Optimization Theory and
  Applications 145~(2) (2010) 387--406.
\newblock \href {http://dx.doi.org/10.1007/s10957-009-9641-1}
  {\path{doi:10.1007/s10957-009-9641-1}}.

\bibitem{Wu2011}
R.~B. Wu, M.~A. Hsieh, H.~Rabitz, Role of controllability in optimizing quantum
  dynamics, Physical Review A 83~(6) (2011) 062306.
\newblock \href {http://dx.doi.org/10.1103/PhysRevA.83.062306}
  {\path{doi:10.1103/PhysRevA.83.062306}}.

\bibitem{Donovan2013}
A.~Donovan, V.~Beltrani, H.~Rabitz, Exploring the impact of constraints in
  quantum optimal control through a kinematic formulation, Chemical Physics 425
  (2013) 46--54.
\newblock \href {http://dx.doi.org/10.1016/j.chemphys.2013.07.019}
  {\path{doi:10.1016/j.chemphys.2013.07.019}}.

\bibitem{Donovan2014a}
A.~Donovan, V.~Beltrani, H.~Rabitz, Local topology at limited resource induced
  suboptimal traps on the quantum control landscape, Journal of Mathematical
  Chemistry 52~(2) (2014) 407--429.
\newblock \href {http://dx.doi.org/10.1007/s10910-013-0269-x}
  {\path{doi:10.1007/s10910-013-0269-x}}.

\bibitem{Donovan2015}
A.~Donovan, H.~Rabitz, Systematically altering the apparent topology of
  constrained quantum control landscapes, Journal of Mathematical Chemistry
  53~(2) (2015) 718--736.
\newblock \href {http://dx.doi.org/10.1007/s10910-014-0453-7}
  {\path{doi:10.1007/s10910-014-0453-7}}.

\bibitem{Riviello2015}
G.~Riviello, K.~M. Tibbetts, C.~Brif, R.~X. Long, R.~B. Wu, T.~S. Ho,
  H.~Rabitz, Searching for quantum optimal controls under severe constraints,
  Physical Review A 91~(4) (2015) 043401.
\newblock \href {http://dx.doi.org/10.1103/PhysRevA.91.043401}
  {\path{doi:10.1103/PhysRevA.91.043401}}.

\bibitem{You2021}
X.~You, X.~Wu, Exponentially many local minima in quantum neural networks
  (2021).
\newblock \href {http://arxiv.org/abs/2110.02479} {\path{arXiv:2110.02479}}.

\bibitem{Wierichs2020}
D.~Wierichs, C.~Gogolin, M.~Kastoryano,
  \href{https://doi.org/10.1103%2Fphysrevresearch.2.043246}{Avoiding local
  minima in variational quantum eigensolvers with the natural gradient
  optimizer}, Physical Review Research 2~(4).
\newblock \href {http://dx.doi.org/10.1103/physrevresearch.2.043246}
  {\path{doi:10.1103/physrevresearch.2.043246}}.
\newline\urlprefix\url{https://doi.org/10.1103%2Fphysrevresearch.2.043246}

\bibitem{Rivera2021}
J.~Rivera-Dean, P.~Huembeli, A.~Ac\'{i}n, J.~Bowles,
  \href{https://arxiv.org/abs/2104.02955}{Avoiding local minima in variational
  quantum algorithms with neural networks} (2021).
\newblock \href {http://dx.doi.org/10.48550/ARXIV.2104.02955}
  {\path{doi:10.48550/ARXIV.2104.02955}}.
\newline\urlprefix\url{https://arxiv.org/abs/2104.02955}

\bibitem{Fontana2020}
E.~Fontana, M.~Cerezo, A.~Arrasmith, I.~Rungger, P.~J. Coles, Optimizing
  parametrized quantum circuits via noise-induced breaking of symmetries
  (2020).
\newblock \href {http://arxiv.org/abs/2011.08763} {\path{arXiv:2011.08763}}.

\bibitem{Lee2021}
J.~Lee, A.~B. Magann, H.~A. Rabitz, C.~Arenz, Towards favorable landscapes in
  quantum combinatorial optimization (2021).
\newblock \href {http://arxiv.org/abs/2105.01114} {\path{arXiv:2105.01114}}.

\bibitem{Anschuetz2021}
E.~R. Anschuetz, \href{https://arxiv.org/abs/2109.06957}{Critical points in
  quantum generative models} (2021).
\newblock \href {http://dx.doi.org/10.48550/ARXIV.2109.06957}
  {\path{doi:10.48550/ARXIV.2109.06957}}.
\newline\urlprefix\url{https://arxiv.org/abs/2109.06957}

\bibitem{Larocca2021}
M.~Larocca, N.~Ju, D.~Garc\'{i}a-Mart\'{i}n, P.~J. Coles, M.~Cerezo,
  \href{https://arxiv.org/abs/2109.11676}{Theory of overparametrization in
  quantum neural networks} (2021).
\newblock \href {http://dx.doi.org/10.48550/ARXIV.2109.11676}
  {\path{doi:10.48550/ARXIV.2109.11676}}.
\newline\urlprefix\url{https://arxiv.org/abs/2109.11676}

\bibitem{Arrasmith2021}
A.~Arrasmith, Z.~Holmes, M.~Cerezo, P.~J. Coles, Equivalence of quantum barren
  plateaus to cost concentration and narrow gorges (2021).
\newblock \href {http://arxiv.org/abs/2104.05868} {\path{arXiv:2104.05868}}.

\bibitem{Mitarai2018}
K.~Mitarai, M.~Negoro, M.~Kitagawa, K.~Fujii, Quantum circuit learning, Phys.
  Rev. A 98 (2018) 032309.
\newblock \href {http://dx.doi.org/10.1103/PhysRevA.98.032309}
  {\path{doi:10.1103/PhysRevA.98.032309}}.

\bibitem{Cerezo2021a}
M.~Cerezo, P.~J. Coles, Higher order derivatives of quantum neural networks
  with barren plateaus, Quantum Science and Technology 6~(3) (2021) 035006.
\newblock \href {http://dx.doi.org/10.1088/2058-9565/abf51a}
  {\path{doi:10.1088/2058-9565/abf51a}}.

\bibitem{Arrasmith2021b}
A.~Arrasmith, M.~Cerezo, P.~Czarnik, L.~Cincio, P.~J. Coles, Effect of barren
  plateaus on gradient-free optimization, {Quantum} 5 (2021) 558.
\newblock \href {http://dx.doi.org/10.22331/q-2021-10-05-558}
  {\path{doi:10.22331/q-2021-10-05-558}}.

\bibitem{Kim2021}
J.~Kim, J.~Kim, D.~Rosa,
  \href{http://dx.doi.org/10.1103/PhysRevResearch.3.023203}{Universal
  effectiveness of high-depth circuits in variational eigenproblems}, Physical
  Review Research 3~(2).
\newblock \href {http://dx.doi.org/10.1103/physrevresearch.3.023203}
  {\path{doi:10.1103/physrevresearch.3.023203}}.
\newline\urlprefix\url{http://dx.doi.org/10.1103/PhysRevResearch.3.023203}

\bibitem{Cerezo2021b}
M.~Cerezo, A.~Sone, T.~Volkoff, L.~Cincio, P.~J. Coles, Cost function dependent
  barren plateaus in shallow parametrized quantum circuits, Nature
  Communications 12~(1) (2021) 1791.
\newblock \href {http://dx.doi.org/10.1038/s41467-021-21728-w}
  {\path{doi:10.1038/s41467-021-21728-w}}.

\bibitem{Sharma2020}
K.~Sharma, M.~Cerezo, L.~Cincio, P.~J. Coles, Trainability of dissipative
  perceptron-based quantum neural networks (2020).
\newblock \href {http://arxiv.org/abs/2005.12458} {\path{arXiv:2005.12458}}.

\bibitem{Patti2021}
T.~L. Patti, K.~Najafi, X.~Gao, S.~F. Yelin, Entanglement devised barren
  plateau mitigation, Physical Review Research 3~(3).
\newblock \href {http://dx.doi.org/10.1103/physrevresearch.3.033090}
  {\path{doi:10.1103/physrevresearch.3.033090}}.

\bibitem{Marrero2021}
C.~O. Marrero, M.~Kieferov\'{a}, N.~Wiebe, Entanglement induced barren plateaus
  (2021).
\newblock \href {http://arxiv.org/abs/2010.15968} {\path{arXiv:2010.15968}}.

\bibitem{Wang2021}
S.~Wang, E.~Fontana, M.~Cerezo, K.~Sharma, A.~Sone, L.~Cincio, P.~J. Coles,
  Noise-induced barren plateaus in variational quantum algorithms (2021).
\newblock \href {http://arxiv.org/abs/2007.14384} {\path{arXiv:2007.14384}}.

\bibitem{Franca2021}
D.~Stilck~Franca, R.~Garcia-Patron, Limitations of optimization algorithms on
  noisy quantum devices, Nature Physics 17~(11) (2021) 1221?1227.
\newblock \href {http://dx.doi.org/10.1038/s41567-021-01356-3}
  {\path{doi:10.1038/s41567-021-01356-3}}.

\bibitem{Arenz2020}
C.~Arenz, H.~Rabitz, Drawing together control landscape and tomography
  principles, Physical Review A 102~(4).
\newblock \href {http://dx.doi.org/10.1103/physreva.102.042207}
  {\path{doi:10.1103/physreva.102.042207}}.

\bibitem{Renes2004}
J.~M. Renes, R.~Blume-Kohout, A.~J. Scott, C.~M. Caves, Symmetric
  informationally complete quantum measurements, Journal of Mathematical
  Physics 45~(6) (2004) 2171--2180.
\newblock \href {http://dx.doi.org/10.1063/1.1737053}
  {\path{doi:10.1063/1.1737053}}.

\bibitem{Harrow2009}
A.~W. Harrow, R.~A. Low, Random quantum circuits are approximate 2-designs,
  Communications in Mathematical Physics 291~(1) (2009) 257--302.
\newblock \href {http://dx.doi.org/10.1007/s00220-009-0873-6}
  {\path{doi:10.1007/s00220-009-0873-6}}.

\bibitem{Holmes2022}
Z.~Holmes, K.~Sharma, M.~Cerezo, P.~J. Coles,
  \href{https://doi.org/10.1103%2Fprxquantum.3.010313}{Connecting ansatz
  expressibility to gradient magnitudes and barren plateaus}, {PRX} Quantum
  3~(1).
\newblock \href {http://dx.doi.org/10.1103/prxquantum.3.010313}
  {\path{doi:10.1103/prxquantum.3.010313}}.
\newline\urlprefix\url{https://doi.org/10.1103%2Fprxquantum.3.010313}

\bibitem{Uvarov2021}
A.~V. Uvarov, J.~D. Biamonte, On barren plateaus and cost function locality in
  variational quantum algorithms, Journal of Physics A: Mathematical and
  Theoretical 54~(24) (2021) 245301.
\newblock \href {http://dx.doi.org/10.1088/1751-8121/abfac7}
  {\path{doi:10.1088/1751-8121/abfac7}}.

\bibitem{Nakaji2021}
K.~Nakaji, N.~Yamamoto, Expressibility of the alternating layered ansatz for
  quantum computation, Quantum 5 (2021) 434.
\newblock \href {http://dx.doi.org/10.22331/q-2021-04-19-434}
  {\path{doi:10.22331/q-2021-04-19-434}}.

\bibitem{Pesah2021}
A.~Pesah, M.~Cerezo, S.~Wang, T.~Volkoff, A.~T. Sornborger, P.~J. Coles,
  Absence of barren plateaus in quantum convolutional neural networks, Physical
  Review X 11~(4).
\newblock \href {http://dx.doi.org/10.1103/physrevx.11.041011}
  {\path{doi:10.1103/physrevx.11.041011}}.

\bibitem{Zhang2020}
K.~Zhang, M.-H. Hsieh, L.~Liu, D.~Tao, Toward trainability of quantum neural
  networks (2020).
\newblock \href {http://arxiv.org/abs/2011.06258} {\path{arXiv:2011.06258}}.

\bibitem{Broers2021}
L.~Broers, L.~Mathey, Optimization of quantum algorithm protocols without
  barren plateaus (2021).
\newblock \href {http://arxiv.org/abs/2111.08085} {\path{arXiv:2111.08085}}.

\bibitem{Larocca2021b}
M.~Larocca, P.~Czarnik, K.~Sharma, G.~Muraleedharan, P.~J. Coles, M.~Cerezo,
  \href{https://arxiv.org/abs/2105.14377}{Diagnosing barren plateaus with tools
  from quantum optimal control} (2021).
\newblock \href {http://dx.doi.org/10.48550/ARXIV.2105.14377}
  {\path{doi:10.48550/ARXIV.2105.14377}}.
\newline\urlprefix\url{https://arxiv.org/abs/2105.14377}

\bibitem{Wiersema2020}
R.~Wiersema, C.~Zhou, Y.~de~Sereville, J.~F. Carrasquilla, Y.~B. Kim, H.~Yuen,
  Exploring entanglement and optimization within the hamiltonian variational
  ansatz, PRX Quantum 1 (2020) 020319.
\newblock \href {http://dx.doi.org/10.1103/PRXQuantum.1.020319}
  {\path{doi:10.1103/PRXQuantum.1.020319}}.

\bibitem{Brandao2018}
F.~G. S.~L. Brandao, M.~Broughton, E.~Farhi, S.~Gutmann, H.~Neven, For fixed
  control parameters the quantum approximate optimization algorithm's objective
  function value concentrates for typical instances (2018).
\newblock \href {http://arxiv.org/abs/1812.04170} {\path{arXiv:1812.04170}}.

\bibitem{Rudolph2021}
M.~S. Rudolph, S.~Sim, A.~Raza, M.~Stechly, J.~R. McClean, E.~R. Anschuetz,
  L.~Serrano, A.~Perdomo-Ortiz, Orqviz: Visualizing high-dimensional landscapes
  in variational quantum algorithms (2021).
\newblock \href {http://arxiv.org/abs/2111.04695} {\path{arXiv:2111.04695}}.

\bibitem{Grant2019}
E.~Grant, L.~Wossnig, M.~Ostaszewski, M.~Benedetti, An initialization strategy
  for addressing barren plateaus in parametrized quantum circuits, {Quantum} 3
  (2019) 214.
\newblock \href {http://dx.doi.org/10.22331/q-2019-12-09-214}
  {\path{doi:10.22331/q-2019-12-09-214}}.

\bibitem{Volkoff2021}
T.~Volkoff, P.~J. Coles, Large gradients via correlation in random
  parameterized quantum circuits, Quantum Science and Technology 6~(2) (2021)
  025008.
\newblock \href {http://dx.doi.org/10.1088/2058-9565/abd891}
  {\path{doi:10.1088/2058-9565/abd891}}.

\bibitem{Skolik2021}
A.~Skolik, J.~R. McClean, M.~Mohseni, P.~van~der Smagt, M.~Leib, Layerwise
  learning for quantum neural networks, Quantum Machine Intelligence 3~(1).
\newblock \href {http://dx.doi.org/10.1007/s42484-020-00036-4}
  {\path{doi:10.1007/s42484-020-00036-4}}.

\bibitem{Verdon2019}
G.~Verdon, M.~Broughton, J.~R. McClean, K.~J. Sung, R.~Babbush, Z.~Jiang,
  H.~Neven, M.~Mohseni, Learning to learn with quantum neural networks via
  classical neural networks (2019).
\newblock \href {http://arxiv.org/abs/1907.05415} {\path{arXiv:1907.05415}}.

\bibitem{Temme2017}
K.~Temme, S.~Bravyi, J.~M. Gambetta,
  \href{https://link.aps.org/doi/10.1103/PhysRevLett.119.180509}{Error
  mitigation for short-depth quantum circuits}, Phys. Rev. Lett. 119 (2017)
  180509.
\newblock \href {http://dx.doi.org/10.1103/PhysRevLett.119.180509}
  {\path{doi:10.1103/PhysRevLett.119.180509}}.
\newline\urlprefix\url{https://link.aps.org/doi/10.1103/PhysRevLett.119.180509}

\bibitem{Endo2021}
S.~Endo, Z.~Cai, S.~C. Benjamin, X.~Yuan,
  \href{http://dx.doi.org/10.7566/JPSJ.90.032001}{Hybrid quantum-classical
  algorithms and quantum error mitigation}, Journal of the Physical Society of
  Japan 90~(3) (2021) 032001.
\newblock \href {http://dx.doi.org/10.7566/jpsj.90.032001}
  {\path{doi:10.7566/jpsj.90.032001}}.
\newline\urlprefix\url{http://dx.doi.org/10.7566/JPSJ.90.032001}

\bibitem{Arute2020}
F.~Arute, K.~Arya, R.~Babbush, D.~Bacon, J.~C. Bardin, R.~Barends, S.~Boixo,
  M.~Broughton, B.~B. Buckley, D.~A. Buell, B.~Burkett, N.~Bushnell, Y.~Chen,
  Z.~Chen, B.~Chiaro, R.~Collins, W.~Courtney, S.~Demura, A.~Dunsworth,
  E.~Farhi, A.~Fowler, B.~Foxen, C.~Gidney, M.~Giustina, R.~Graff, S.~Habegger,
  M.~P. Harrigan, A.~Ho, S.~Hong, T.~Huang, W.~J. Huggins, L.~Ioffe, S.~V.
  Isakov, E.~Jeffrey, Z.~Jiang, C.~Jones, D.~Kafri, K.~Kechedzhi, J.~Kelly,
  S.~Kim, P.~V. Klimov, A.~Korotkov, F.~Kostritsa, D.~Landhuis, P.~Laptev,
  M.~Lindmark, E.~Lucero, O.~Martin, J.~M. Martinis, J.~R. McClean, M.~McEwen,
  A.~Megrant, X.~Mi, M.~Mohseni, W.~Mruczkiewicz, J.~Mutus, O.~Naaman,
  M.~Neeley, C.~Neill, H.~Neven, M.~Y. Niu, T.~E. O’Brien, E.~Ostby,
  A.~Petukhov, H.~Putterman, C.~Quintana, P.~Roushan, N.~C. Rubin, D.~Sank,
  K.~J. Satzinger, V.~Smelyanskiy, D.~Strain, K.~J. Sung, M.~Szalay, T.~Y.
  Takeshita, A.~Vainsencher, T.~White, N.~Wiebe, Z.~J. Yao, P.~Yeh, A.~Zalcman,
  Hartree-fock on a superconducting qubit quantum computer, Science 369~(6507)
  (2020) 1084--1089.
\newblock \href {http://dx.doi.org/10.1126/science.abb9811}
  {\path{doi:10.1126/science.abb9811}}.

\bibitem{Wu2019}
R.~B. Wu, Q.~Y. Sun, T.~S. Ho, H.~Rabitz, Inherently trap-free convex
  landscapes for fully quantum optimal control, Journal of Mathematical
  Chemistry 57~(9) (2019) 2154--2167.
\newblock \href {http://dx.doi.org/10.1007/s10910-019-01059-4}
  {\path{doi:10.1007/s10910-019-01059-4}}.

\bibitem{Banchi2020}
L.~Banchi, J.~Pereira, S.~Lloyd, S.~Pirandola, Convex optimization of
  programmable quantum computers, npj Quantum Information 6~(1) (2020) 42.
\newblock \href {http://dx.doi.org/10.1038/s41534-020-0268-2}
  {\path{doi:10.1038/s41534-020-0268-2}}.

\end{thebibliography}
\end{document}